\documentstyle[12pt]{article}
\newcommand{\bq}{\begin{equation}}
\newcommand{\eq}{\end{equation}}
\newcommand{\bqn}{\begin{eqnarray}}
\newcommand{\eqn}{\end{eqnarray}}
\newcommand{\nb}{\nonumber}
\newcommand{\lb}{\label}

\newcommand{\bb}[1]{\begin{equation}\label{#1}}
\newcommand{\ee}{\end{equation}}
\newcommand{\bbb}{\begin{eqnarray}}
\newcommand{\eee}{\end{eqnarray}}
\newcommand{\bbbb}{\begin{eqnarray*}}
\newcommand{\eeee}{\end{eqnarray*}}

\begin{document}

\title{Gravitational Collapse of Phantom Fluid
in $(2+1)$-Dimensions\footnote{modified version}}
\author{M.\,R. Martins$^1$, M.\,F.\,A. da Silva$^1$
and  A.\,D. Sheng$^2$ \\
\small\it $^1$ Departamento de F\'{\i}sica Te\' orica,
Universidade do Estado do Rio de Janeiro, \\
\small\it Rua S\~ao Francisco Xavier $524$, Maracan\~ a,
$20550-013$ \\
\small \it Rio de Janeiro, RJ, Brazil\\
\small \it $^2$ Department of Computer Science, Carnegie Mellon
University\\ \small \it  SMC 2075,
Pittsburgh, PA 15289-2075, USA\\
\small \it E-mail Addresses: mrmar@dft.if.uerj.br; mfas@dtf.if.uerj.br; \\
\small \it asheng@andrew.cmu.edu}


\maketitle

\begin{abstract}
\noindent This paper is devoted to the solutions of Einstein's field
equations for a circularly symmetric anisotropic fluid, with
kinematic self-similarity of the first kind, in $(2+1)$-dimensional
spacetimes. In the case where the radial pressure vanishes, we show
that there exists a solution of the equations that represents the
gravitational collapse of an anisotropic fluid, and this collapse
will eventually form a black hole, even when it is constituted by
the phantom energy.
\end{abstract}

\section{Introduction}

Self-similar solutions of Einstein field equations have attracted a
great deal of recent attentions, not only because they can be
studied analytically through simplifications, but also due to their
relevance in astrophysics \cite{Car} and critical phenomena in
gravitational collapses \cite{Chop93,SS1}.

Lately, we investigated the self-similar solutions in various
spacetimes \cite{SS}. In particular, we considered a massless scalar
field in $(2+1)$-dimensional circularly symmetric spacetimes with
kinematic self-similarity of the second kind in the context of
Einstein's theory, and acquired all such solutions \cite{fatima05}.
We further studied their local and global properties and found that
some of them represent gravitational collapses of a massless scalar
field, in which black holes are always formed. In another discussion
we discussed an anisotropic fluid, with the same self-similarity,
and showed that the existing solution is unique when the radial
pressure vanishes, and it represents a collapsing dust fluid, where
the final output can be either naked singularities or black holes
\cite{Marina}.

In this paper, we extend the aforementioned studies to the case of
an anisotropic fluid with zero radial pressure while the
self-similarity of the first kind is considered.

\bigskip

\section{Solutions of the field equations}

The general metric can be given by

\smallskip

\bq
\label{1}
ds^{2}=e^{2\phi }dt^{2}-e^{2\psi }dr^{2}-r^{2}S^{2}d\theta^{2}.
\eq

\smallskip

The non-null components of Einstein's tensor are

\smallskip

\begin{eqnarray}
\label{2} G_{tt}&=&\frac{e^{-2\psi }}{r\,S}\left\{ e^{2\phi }\psi
_{,r}rS_{,r}+e^{2\phi }\psi _{,r}S-e^{2\phi }rS_{,rr\,}-2e^{2\phi
\,}S_{,r\,}+e^{2\psi }\psi _{,t\,}rS_{,t}\right\},~~\\
\label{3}
G_{tr}&=&\frac{1}{r\,S}\left\{ \phi _{,r}rS_{,t\,}+\psi
_{,t}rS_{,r}+\psi _{,t}S-rS_{,tr}-S_{,t}\right\},\\
\label{4} G_{rr}&=&\frac{e^{-2\phi }}{r\,S}\left\{ e^{2\phi }\phi
_{,r}rS_{,r\,}+e^{2\phi }\phi _{,r}S+e^{2\psi }\phi
_{,t}rS_{,t}-e^{2\psi }rS_{,tt}\right\} , \\
\label{5} G_{\theta \theta } &=& r^{2}S^{2}\left\{e^{-2\psi }\phi
_{,r}^{2}-e^{-2\psi }\phi _{,r}\psi _{,r}+e^{-2\psi }\phi
_{,rr}+e^{-2\phi }\phi _{,t}\psi _{,t}\right.
\nonumber \\
&&\left. -e^{-2\phi }\psi _{,t}^{2}-e^{-2\phi }\psi _{,tt}\right\}.
\end{eqnarray}

\smallskip

To study properties of these solutions with self-similarity of the
first kind, we introduce two new dimensionless variables, \,$\chi$\,
and \,$\tau,$ through the relations

\smallskip

\begin{eqnarray}
\label{6} \chi&=&\ln\left(z\right)=\ln\left[\frac{r}{(-t)}\right],
\\
\label{7} \tau &=&-\ln\left( -t\right).
\end{eqnarray}

\smallskip

Note that the range of the time coordinate considered is
$-\infty<t\leq 0$.

On the other hand, a self-similar solution is defined by \bq
\lb{8.a} \phi(t,r) = \phi(\chi),\;\;\; \psi(t,r) = \psi(\chi),\;\;\;
S(t,r) = S(\chi). \eq Thus,  Eqs. (\ref{2})-(\ref{5}) become,

\smallskip

\begin{eqnarray}
\label{8} G_{tt}&=&-\frac{1}{r^{2}Se^{2\psi }}\left\{e^{2\phi
}\left[ S_{,\chi \chi }+S_{,\chi }-\psi _{,\chi }\left( S_{,\chi
}+S\right) \right] -\frac{r^{2}}{t^{2}}\psi _{,\chi }S_{,\chi
}e^{2\psi }\right\} , \\
\label{9} G_{tr}&=&\frac{1}{tr\,S}\left\{ S_{,\chi \chi \,} -\psi
_{,\chi}\left( S_{,\chi }+S\right) -S_{,\chi }\left( \phi _{,\chi
}-1\right) \right\} ,\\
\nb G_{rr} & = & \frac{1}{r^{2\,}S\,e^{2\phi }}\left\{e^{2\phi
}\left[ \phi _{,\chi }\left( S_{,\chi }+S\right) \right]\rule{0mm}{6mm}\right.\\
\label{10}
&&  \left.-\frac{%
r^{2}\,e^{2\psi }}{t^{2}}(S_{,\chi \chi }-S_{,\chi \,} \phi _{,\chi
})-\frac{r^{2}}{t^{2}}e^{2\psi }S_{,\chi }\right\} , \\
G_{\theta
\theta } &=& S^{2}\left\{e^{-2\psi }\left[ \phi _{,\chi \chi }+\phi
_{,\chi }\left( \phi _{,\chi }-\psi _{,\chi }-1\right) \right]
-\frac{r^{2}e^{-2\phi }}{t^{2}}[\psi
_{,\chi \chi }\right.  \nonumber \\
&& \left. -\psi _{,\chi }\left( \phi _{,\chi }-\psi _{,\chi
}-1\right) ]\rule{0mm}{6mm}\right\}.\label{11}
\end{eqnarray}

\smallskip

The momentum-energy tensor is given by

\smallskip

\begin{equation}
\label{12}
T_{\mu \nu }\,=\,\rho \,u_{\mu }\,u_{\nu }\,+\,p_{r}\,r_{\mu }\,r_{\nu
}+p_{\theta }\,\theta _{\mu }\theta _{\nu },
\end{equation}

\smallskip

\noindent where \,$\rho$\, is the energy density, \,$p_{r}$ and
$p_{\theta }$\, are the radial and the tangential pressures,
respectively, with \,$u_{\mu }$\,, \,$r_{\mu}$\, and \,$\theta
_{\mu}$\, being given by

\begin{equation}
\label{13} u_{\mu }=e^{\phi \left( \chi \right) }\delta _{\mu
}^{t},\,\ \ \ \,\,\,\,r_{\mu }\,=\,e^{\psi \left( \chi \right)
}\delta _{\mu }^{r},\,\,\,\,\,\,\,\,\theta _{\mu }\,=\,r\,S\left(
\chi \right) \,\delta _{\mu }^{\theta },
\end{equation}

\smallskip

\noindent where a comoving frame is adopted.

Substituting Eqs. (\ref{8})-(12) and (\ref{12}) into the following
field equations

\smallskip

\begin{equation}
\label{14}
G_{\mu \nu }\,=\kappa T_{\mu \nu },
\end{equation}

\smallskip

\noindent we obtain that

\smallskip

\begin{eqnarray}
\label{15} \rho &=&-\frac{1}{\kappa}\left\{\frac{1}{r^{2}}e^{-2\psi
}\left[ \,y_{,\chi \,}+(\,y+1)(\,y-\psi _{,\chi })\right]
-\frac{1}{t^{2}}
e^{-2\phi }\psi _{,\chi }\,y\right\},\\
\label{16} y_{,\chi \,}&=&\,y\phi _{,\chi }+(\,y+1)(\psi _{,\chi
}-y)\,,\\
\label{17}
p_{r}&=&\frac{1}{l^{2}\kappa}\left\{\frac{1}{r^{2\,}}e^{-2\psi }\phi
_{,\chi }(\,y+1)-\frac{1}{t^{2}}\,e^{-2\phi }\left[ \,y_{,\chi
\,}+\,y(\,y-\phi _{,\chi }+1)\right] \right\},\\
p_{\theta } &=&\frac{1}{l^{2}\kappa}\left\{\frac{e^{-2\psi
}}{r^{2\,}}\left[ \phi
_{,\chi \chi \,}+\phi _{,\chi }(\phi _{,\chi }-\psi _{,\chi }-1)\right] -%
\frac{e^{-2\phi }}{t^{2}}[\psi _{,\chi \chi \,}+\right.  \nonumber \\
&&\left.-\psi _{,\chi }(\phi _{,\chi }-\psi _{,\chi
}-1)]\rule{0mm}{6mm}\right\}.\label{18}
\end{eqnarray}

\smallskip

Note that in above equations we set

\smallskip

\begin{equation}
\label{19} y\left( \chi \right) =\frac{S_{,\chi }}{S}~.
\end{equation}

\smallskip

Therefore we have 4 equations for the 6 functions to determine, that
is, $\phi,~\psi,~S,~\rho,~p_{r}$ and $p_{\theta }.$ For this, we
introduce two additional equations, $p_{r}=0$ and $p_{\theta
}=\omega\rho$ for forming a consistent and solvable system. The two
additional state equations furnish

\smallskip

\begin{equation}
\label{20}
\phi _{,\chi }(\,y+1)-e^{2(\chi-\phi+\psi)}\left[ \,y_{,\chi }+\,y(\,y-\phi
_{,\chi }+1 )\right]=0
 \end{equation}

\smallskip

\noindent and

\smallskip

\begin{eqnarray}
\label{21}
&& \phi_{,\chi \chi \,}+\phi _{,\chi }(\phi _{,\chi }-\psi _{,\chi }-1)\nonumber\\
&&~~~+e^{2(\chi-\phi+\psi)}[-\omega\psi,_\chi y-\psi _{,\chi \chi \,}+\psi _{,\chi }(\phi _{,\chi }-\psi _{,\chi }-1)]\nonumber\\
&&~~~+\omega\left[y_{,\chi \,}+(\,y+1)(\,y-\psi _{,\chi })\right]=0
\end{eqnarray}

\smallskip

A substitution of equation (\ref{16}) into equation (\ref{20})
yields

\smallskip

\begin{equation}
\label{22}
(y+1)\left[\phi_{,\chi}-e^{2(\chi-\phi+\psi)}\psi_{,\chi} \right]=0.
\end{equation}

\smallskip

Thus, we have two possible solution equations which are

\smallskip

\begin{eqnarray}
\label{23} (i) &&y+1=0, \\
\label{24} (ii) &&\phi_{,\chi}-e^{2(\chi-\phi+\psi)}\psi_{,\chi}=0.
\end{eqnarray}

\smallskip

Although we are not able to solve analytically equation (\ref{24}),
for the former case $(i),$ we have the solutions

\smallskip

\begin{eqnarray}
\label{25} S\left( \chi \right) &=&S_0 e^{-\chi},\\
\label{26} \psi\left( \chi \right) &=&\psi
_{0}+\ln\left[a+e^{(\omega-1)\chi}\right],
\end{eqnarray}

\smallskip

\noindent and

\smallskip

\begin{equation}
\label{27} \phi\left( \chi \right) ~=~\phi _{0},
\end{equation}

\smallskip

\noindent where $\phi_0,~\psi_0,~S_0$ and $a$ are integration
constants.

Thus, without lost of a generality, the corresponding metric
(\ref{1}) can be reformulated in the form,

\smallskip

\begin{eqnarray}
\label{met1}
{ds}^2 &=&dt^2-\left[a+\left(\frac{r}{-t}\right)^{(\omega-1)}\right]^2
dr^2-S_0^2 t^2d\theta^2,
\end{eqnarray}

\smallskip

\noindent due to the fact that we have set $\phi_0=\psi_0=0.$ The
energy density and the tangential pressure can be written,
respectively, as

\smallskip

\begin{equation}
\label{29}
\rho \left( r,t\right) =\frac{(1-\omega)r^{(\omega-1)}}{l^2 K t^2
\left[(-t)^{(\omega-1)}a+r^{(\omega-1)}\right]}
\end{equation}

\smallskip

\noindent and

\smallskip

\begin{equation}
\label{30}
p_\theta \left( r,t\right) = \frac{\omega(1-\omega)r^{(\omega-1)}}{l^2 K t^2
\left[(-t)^{(\omega-1)}a+r^{(\omega-1)}\right]}.
\end{equation}

\smallskip

The Kretschmann's scalar, for above solutions are given by

\smallskip

\begin{equation}
\label{31}
K\left( r,t\right) =\frac{4(\omega-1)^2(\omega^2+1)}{t^4\,e^{4\phi_0}
\left[1+ar^{(1-\omega)}(-t)^{(\omega-1)}\right]^2}.
\end{equation}

\smallskip

So, there is the possibility of the formation of two singularities
which are

\smallskip

\begin{equation}
\label{32}
t=t_{sing}=0
\end{equation}

\smallskip

\noindent and

\smallskip

\begin{equation}
\label{33} r=r_{sing}=\left[
\frac{(-t)}{|\,a|^{1/(1-\omega)}}\right],\quad {\rm for}\quad a < 0.
\end{equation}

\smallskip

Further, the geometric radius, $Rg,$ is defined by

\smallskip

\begin{equation}
\label{34}
Rg=S\;r=-S_{0\,}t,
\end{equation}

\smallskip

\noindent where $S_{0}\geq 0.$

We note the fact that a geometric radius decreases with respect to
the time indicates that the process represents a collapse.

The expansions of the ingoing and outgoing null geodesics congruence
is useful for understanding global properties of the solutions. The
latter are given by

\smallskip

\begin{equation}
\label{35}
\theta_{l}=\theta_{n}=-S_0.
\end{equation}

\smallskip

Now, observe that both $\theta_{l}$ and $\theta_{n}$ are always
negative. Therefore, the solutions we have obtained must represent a
region inside the event horizon of a black hole. To support this
hypothesis, it is necessary to cut the current spacetime and match
it with a different spacetime which represents the exterior of a
black hole.

\bigskip

\section{The energy conditions}

For $a>0,$ the signs of the energy density and of the tangential
pressure can be determined by the terms $1-\omega$ and
$\omega(1-\omega),$ respectively. Thus, it is clear to see that for
$-1\leq \omega\leq 1,$ all the required energy conditions are
satisfied. If we admit the dark energy, we can construct a collapse
model of a phantom anisotropic fluid, with $\omega\leq -1.$ In the
case, all the energy conditions are violated, while the energy
density preserves its positivity.

\bigskip

\section{The junction conditions}

\subsection{Junction conditions without thin shells}

\smallskip

In order to match the interior fluid solution with the exterior
vacuum solution we consider the Darmois junction conditions
\cite{Dar}, which requires that the first and the second fundamental
forms (that are the metric and the extrinsic curvature,
respectively) be continuous across the junction hypersurface.

We divide the current spacetime into following two regions: one of
them is called the interior region, constituted by a circularly
symmetric anisotropic fluid with kinematic self-similarity of the
first kind (\,$V^-$\,, for \,$r\leq r_\Sigma$\,), and the other,
called the exterior region, fulfilled by vacuum in the presence of
negative cosmological constant (\,$V^+$\,, em \,$r\geq r_\Sigma$\,),
where $r_\Sigma$ is the radial coordinate of the hypersurface.

The general metric which describes the internal region, $ds_-^{2},$
has been given by equation (\ref{met1}), with $0\leq r\leq
r_\Sigma,~0\leq \theta < 2\pi$ and $-\infty < t \leq 0.$

As for the exterior spacetime, we have the BTZ solution described by
the metric

\smallskip

\bq \label{36} ds_+^{2}=-(\Lambda R^2+M)dT^2+\frac{1}{\Lambda
R^2+M}dR^2-R^2d\theta^2, \eq

\smallskip

\noindent where $R_\Sigma\leq R < \infty,~ -\infty < T \leq 0,~ 0
\leq \theta <2\pi,~ \Lambda < 0$ and $M,$ the mass of the black
hole, where event horizon is located on
$$R_{EH}=\sqrt{\frac{-M}{\Lambda}}~.$$

On the junction hypersurface, the intrinsic metric reduces to

\smallskip

\bq \label{37} ds^2_{\Sigma}=d\tau^2-H(\tau)^2_{\Sigma}\,d\theta^2,
\eq

\smallskip

\noindent where $\tau$ is the proper time.

To apply the junction conditions, we first require the continuity of
the metric potential, that is,

\smallskip

\bq \label{38}
\left(ds_{-}^{2}\right)_{\Sigma}=\left(ds^{2}\right)_{\Sigma}=
\left(ds_{+}^{2}\right)_{\Sigma}, \eq

\smallskip

\noindent which indicates that

\smallskip

\begin{eqnarray}
\label{39} d\tau&=&dt, \\
\label{40} \frac{dt}{dT}&=&\sqrt{\frac{-(\Lambda
R_{\Sigma}^{2}+M)^{2}+\biggl(\frac{dR_{\Sigma}(T)}{dT}\biggr)^2}{(\Lambda
R_{\Sigma}^{2}+M)}}, \\
\label{41} R_{\Sigma}^{2}(T)&=&S_0^{2}\tau^2.
\end{eqnarray}

\smallskip

Further, the second fundamental form, or the extrinsic curvature, is
defined as

\smallskip

\bq \label{42}
K_{ij}^{\pm}=\eta_{\alpha}^{\pm}\frac{\partial^2x_{\pm}^{\alpha}}
{\partial\xi^i\partial\xi^j}-\eta_{\alpha}^{\pm}\Gamma_{\mu\nu}^{\alpha}
\frac{\partial x_{\pm}^{\mu}}{\partial\xi^i}\frac{\partial
x_{\pm}^{\nu}}{\partial\xi^j}\,, \eq

\smallskip

\noindent where

\smallskip

\bq \label{43} \eta_{\alpha}^{\pm}=L\frac{\partial f^{\pm}}{\partial
x^{\alpha}}\,, \eq

\smallskip

\noindent are the unit normal vectors and \,$f^{\pm}$\, is the
function which defines the hypersurface, given by

\smallskip

\begin{eqnarray}
\label{44} f^-&=&r-r_{\Sigma}=0, \\
\label{45} f^+&=&R-R_{\Sigma}(T)=0.
\end{eqnarray}

\smallskip

On the other hand, the unit normal vectors,
\,$\eta^{\pm}_{\alpha},$\, are given respectively by

\smallskip

\bq \label{46}
\eta^{-}_{\alpha}=\left[a+\biggl(\frac{r_{\Sigma}}{-t}\biggr)^{(\omega-1)}\right](0,1,0),
\eq

\smallskip

\noindent and

\smallskip

\bq \label{47} \eta^{+}_{\alpha}=\sqrt{\frac{\Lambda
R_{\Sigma}^2+M}{-(\Lambda
R_{\Sigma}^2+M)^2+\biggl(\frac{dR_{\Sigma}(T)}{dT}\biggr)^2}}
\left(\frac{-dR_{\Sigma}(T)}{dT},1,0\right). \eq

\bigskip

We find that all components of the extrinsic curvature for the
interior spacetime are null, while for the exterior spacetime the
non-null components are only given by

\smallskip

\begin{eqnarray}
K_{00}^{+}&=&\sqrt{\frac{\Lambda R_{\Sigma}^2+M}{-(\Lambda
R_{\Sigma}^2+M)^2
+\biggl(\frac{\partial R_{\Sigma}(T)}{\partial T}\biggr)^2}}\nonumber\\
&&\times\left\{\frac{\partial^{2}R_{\Sigma}(T)}{\partial\tau^2}
+3\left(\frac{\partial R_{\Sigma}(T)}{\partial\tau}\right)^2
\left(\frac{\Lambda R_{\Sigma}}{\Lambda R_{\Sigma}^2+M}\right)\nonumber\right.\\
&&\left.-\Lambda R_{\Sigma}(\Lambda
R_{\Sigma}^2+M)\left(\frac{\partial
T}{\partial\tau}\right)^2-\frac{\partial R_{\Sigma}(T)}{\partial
T}\frac{\partial^2 T}{\partial\tau^2}\right\}\label{48}
\end{eqnarray}

\smallskip

\noindent as well as

\smallskip

\bq \label{49} K_{22}^{+}=-\sqrt{\frac{\Lambda
R_{\Sigma}^2+M}{-(\Lambda R_{\Sigma}^2+M)^2+\biggl(\frac{\partial
R_{\Sigma}(T)}{\partial T}\biggr)^2}}~R_{\Sigma}\biggl(\Lambda
R_{\Sigma}^2+M\biggr). \eq

\smallskip

Then the continuity of the second fundamental forms, which are given
by

\smallskip

\bq \label{50}
\left(K_{00}^{-}\right)_{\Sigma}=\left(K_{00}\right)_{\Sigma}=
\left(K_{00}^{+}\right)_{\Sigma} \eq

\smallskip

\noindent and

\smallskip

\bq \label{51}
\left(K_{22}^{-}\right)_{\Sigma}=\left(K_{22}\right)_{\Sigma}=
\left(K_{22}^{+}\right)_{\Sigma}, \eq

\smallskip

\noindent furnishes

\smallskip

\begin{eqnarray}
&&\sqrt{\frac{\Lambda R_{\Sigma}^{2}+M}{-(\Lambda
R_{\Sigma}^{2}+M)^2+
\biggl(\frac{\partial R_{\Sigma}(T)}{\partial T}\biggr)^2}}\nonumber\\
&&~~~~~\times\left\{\frac{\partial^2R_{\Sigma}(T)}{\partial\tau^2}
+3\left(\frac{\partial R_{\Sigma}(T)}{\partial\tau}\right)^2
\left(\frac{\Lambda R_{\Sigma}}{\Lambda R_{\Sigma}^{2}+M}\right)\right.\nonumber\\
&&~~~~~\left.-\Lambda R_{\Sigma}(\Lambda
R_{\Sigma}^{2}+M)\left(\frac{\partial
T}{\partial\tau}\right)^2-\frac{\partial R_{\Sigma}(T)}{\partial
T}\frac{\partial^2 T}{\partial\tau^2}\right\}=0\label{52}
\end{eqnarray}

\smallskip

\noindent and

\smallskip

\bq \label{53} -\sqrt{\frac{\Lambda R_{\Sigma}^2+M}{-(\Lambda
R_{\Sigma}^2+M)^2+\biggl(\frac{\partial R_{\Sigma}(T)}{\partial
T}\biggr)^2}}~R_{\Sigma}\biggl(\Lambda R_{\Sigma}^2+M\biggr)=0. \eq

\smallskip

It follows that there are two solutions, which satisfy the
constraints given by equations (\ref{52}) and (\ref{53})
simultaneously, and they are

\smallskip

\bq \label{54} R_{\Sigma}(T)=0 \eq

\smallskip

\noindent and

\smallskip

\bq \label{55} R_{\Sigma}(T)=\sqrt{\frac{-M}{\Lambda}}=R_{EH}. \eq

\smallskip

\noindent Equations (\ref{54}) and (\ref{55}) imply that it is not
possible to matching the spacetimes without the introduction of a
thin shell.

\bigskip

\subsection{Junction conditions with thin shells}

It is observed that conditions (\ref{39}) and (\ref{40}) furnish

\smallskip

\bq \label{56} \frac{dT}{d\tau}=\sqrt{\frac{S_{0}^2-(\Lambda
R_{\Sigma}^2+M)}{(\Lambda R_{\Sigma}^2+M)^2}}, \eq

\smallskip

\noindent and

\smallskip

\bq \label{57} \frac{d^2T}{d\tau^2}=\frac{S_0 \Lambda
R_{\Sigma}}{(\Lambda R_{\Sigma}^2+M)^2 \sqrt{{S_0}^2-(\Lambda
R_{\Sigma}^2+M)}}\left[2{S_0}^2-(\Lambda R_{\Sigma}^2+M)\right]. \eq

\smallskip

Substituting the equations (\ref{56}), (\ref{57}) into (\ref{48})
and (\ref{49}), we may rewrite the components of the $K_{ij}^{+}$
tensor as

\smallskip

\bq \label{58} K_{00}^{+}=\frac{\Lambda
R_{\Sigma}\left\{4S_{0}^{4}-\left(\Lambda
R_{\Sigma}^2+M\right)\left[2S_{0}^{2}+\left(\Lambda
R_{\Sigma}^2+M\right)\right]\right\}}{\left(\Lambda
R_{\Sigma}^2+M\right)^2\sqrt{S_{0}^{2}-\left(\Lambda
R_{\Sigma}^2+M\right)}}, \eq

\smallskip

\noindent and

\smallskip

\bq \label{59} K_{22}^{+}=-R\sqrt{{S_0}^2-(\Lambda R_{\Sigma}^2+M)}.
\eq

\smallskip

The momentum-energy tensor of the thin shell can be rewritten in an alternative form as

\smallskip

\bq \label{60} \Pi_{\mu\nu}=\frac{1}{\kappa}
\left\{K_{\mu\nu}^{-}-K_{\mu\nu}^{+}-g_{\mu\nu}\,g^{\alpha\beta}\left[K_{\alpha\beta}^{-}
-K_{\alpha\beta}^{+}\right]\right\},
\eq

\smallskip

\noindent where $\kappa=8\pi.$ The non-null components are thus
given by

\smallskip

\bq \label{61} \Pi_{00}=-\frac{1}{R_{\Sigma}\kappa}\left\{(\Lambda
R_{\Sigma}^2+M)\sqrt{S_{0}^{2}-\left(\Lambda
R_{\Sigma}^2+M\right)}\right\} \eq


\smallskip

\noindent and

\smallskip

\bq \label{62} \Pi_{22}=\frac{\Lambda R_{\Sigma}^3}{\kappa}
\left\{\frac{4S_{0}^{4}-(\Lambda
R_{\Sigma}^2+M)\left[2S_{0}^{2}+(\Lambda
R_{\Sigma}^2+M)\right]}{(\Lambda
R_{\Sigma}^2+M)^3\sqrt{S_{0}^{2}-(\Lambda R_{\Sigma}^2+M)}}\right\}.
\eq

\smallskip

We can also rewrite the $\Pi_{\mu\nu}$ tensor in the form

\smallskip

\bq
\label{63}
\Pi_{\mu\nu}=\sigma\,u_{\mu}u_{\nu}+\xi_\theta\,\Theta_{\mu}\Theta_{\nu},
\eq

\smallskip

\noindent where $\sigma$ and $\xi$ are the energy density and the
tangential pressure of the shell, respectively, and
$$u_{\mu}=\sqrt{g_{oo}}~\delta_{\mu}^{t}~~\mbox{and}~~
\Theta_{\mu}=\sqrt{-g_{22}}~\delta_{\mu}^{\theta}.$$

\smallskip

Utilizing the components of the metric, momentum-energy tensor and
equations (\ref{61}) and (\ref{62}) for equation (\ref{63}), we
observe immediately that the shell's energy density and the
tangential pressures can be given by

\smallskip

\bq \label{64}
\sigma=\frac{1}{R_{\Sigma}\kappa}\sqrt{S_{0}^{2}-(\Lambda
R_{\Sigma}^2+M)} \eq

\smallskip

\noindent and

\smallskip

\bq \label{65} \xi_{\theta}=\frac{\Lambda R_{\Sigma}}{\kappa}
\left\{\frac{4S_{0}^{4}-\left(\Lambda
R_{\Sigma}^2+M\right)\left[2S_{0}^{2}+(\Lambda
R_{\Sigma}^2+M)\right]}{(\Lambda
R_{\Sigma}^2+M)^3\sqrt{S_{0}^{2}-(\Lambda R_{\Sigma}^2+M)}}\right\}.
\eq

\bigskip

\subsection{The energy conditions of thin shell}

A weak energy condition requires

\smallskip

\bq
\label{66}
\sigma\geq 0
\eq

\smallskip

\noindent and

\smallskip

\bq
\label{67}
\sigma+\xi_\theta\geq 0,
\eq

\smallskip

\noindent while the strong energy condition requires only the
condition (\ref{67}) and the dominant energy condition requires
equations (\ref{66}), (\ref{67}) and further,

\smallskip

\bq
\label{68}
\sigma-\xi_\theta\geq 0.
\eq

\smallskip

In order to secure real $\sigma,$ we must have

\smallskip

\bq \label{69} S_{0}^2-(\Lambda R_{\Sigma}^2+M)\geq 0~~\rightarrow~~
S_{0}^2\geq \Lambda R_{\Sigma}^2+M, \eq

\smallskip

\noindent which implies that the value of $\sigma$ is always
positive, while equations (\ref{67}) and (\ref{68}) furnish,
respectively,

\smallskip

\begin{eqnarray}\label{70} S_0^2-(\Lambda R_{\Sigma}^2+M)&\geq&
\frac{\mid\Lambda\mid R_{\Sigma}\,F}{(\Lambda R_{\Sigma}^2+M)^3},
\\
\label{71} S_0^2-(\Lambda R_{\Sigma}^2+M)&\geq&
-\frac{\mid\Lambda\mid R_{\Sigma}\,F}{(\Lambda R_{\Sigma}^2+M)^3},
\end{eqnarray}

\smallskip

\noindent where $F=-\Lambda^2 R_{\Sigma}^4-2\Lambda(M+S_0^2)R_{\Sigma}^2+G(S_0,M).$

\smallskip

The metric (\ref{36}), which describes the BTZ solution, is defined
only for $(\Lambda R_{\Sigma}^2+M)<0,$ representing the exterior
spacetime of a black hole, which indicates that $S_0^2-(\Lambda
R_{\Sigma}^2+M)\geq 0$ is always true. Thus, to summarize, we may
rewrite conditions (\ref{70}) and (\ref{71}) for the cases when
$F<0$ and $F\geq 0$ as

\smallskip

\begin{eqnarray}
(i)&& F<0:\nonumber\\
\label{72} &&S_0^2-(\Lambda R_{\Sigma}^2+M)\geq\frac{\mid\Lambda\mid
R_{\Sigma} \mid F\mid}{\mid\Lambda R_{\Sigma}^2+M\mid^3}\\
&&~~~~~\mbox{for}~~S_0^2-(\Lambda
R_{\Sigma}^2+M)<-\left[S_0^2+S_0\sqrt{S_0^2+4}\,\right],\nonumber\\
(ii)&& F\geq 0:\nonumber\\
\label{73} &&S_0^2-(\Lambda R_{\Sigma}^2+M)\geq\frac{\mid\Lambda\mid
R_{\Sigma}\,F}{\mid\Lambda R_{\Sigma}^2+M\mid^3}\\
&&~~~~~\mbox{for}~~-\left[S_0^2+S_0\sqrt{S_0^2+4}\,\right]\leq
S_0^2-(\Lambda R_{\Sigma}^2+M)<0. \nonumber\end{eqnarray}

\bigskip

\section{Conclusion}

We obtain a solution of the Einstein equations for an anisotropic
and circularly symmetric, self-similar fluid of the first kind in a
(2+1)-dimensional spacetime. We introduce the state equations,
$p_{r}=0$ as well as $p_{\theta}=\omega\rho,$ in order to solve the
anticipated problem. It is shown that there is such a solution which
represents gravitational collapse of an anosotropic fluid. The final
output can be either a normal black hole or a black hole made of
phantom.

\bigskip

\section*{Acknowledgments}

The partial financial supports from UERJ (MRM), FAPERJ/UERJ, CNPq
(MFAdaS) and Baylor University (URSA Grant No: 030150613) are
gratefully acknowledged.


\end{document}